\RequirePackage{fix-cm}
\documentclass[reqno]{amsproc}
\usepackage{fixltx2e}
\usepackage{amssymb}

\usepackage[citation-order,nobysame]{amsrefs}
\usepackage{xyzbib}

\usepackage{hyperref}

\usepackage{pstricks,pst-node,pst-plot}

\numberwithin{equation}{section}
\let\cite=\cites

\newcommand\vc{v_\mathrm{c}}
\newcommand\Rc{R_\mathrm{c}}

\newcommand{\rme}{\mathrm{e}}
\newcommand{\rmi}{\mathrm{i}}
\newcommand{\rmd}{\mathrm{d}}

\begin{document}
\title{Third-order phase transition in random tilings}

\author{F. Colomo}
\address{INFN, Sezione di Firenze\\
Via G. Sansone 1, 50019 Sesto Fiorentino (FI), Italy}
\email{colomo@fi.infn.it}

\author{A. G. Pronko}
\address{St.~Petersburg Department of V.~A.~Steklov Mathematical Institute of the
Russian Academy of Sciences,
Fontanka 27, 191023 St.~Petersburg, Russia}
\email{agp@pdmi.ras.ru}

\begin{abstract}
We consider the domino tilings of an Aztec diamond with a cut-off corner of
macroscopic square shape and given size, and address the bulk properties of
tilings as the size is varied.  We observe that the free energy exhibits a
third-order phase transition when the cut-off square, increasing in size,
reaches the arctic ellipse---the phase separation curve of the original
(unmodified) Aztec diamond.  We obtain this result by studying the
thermodynamic limit of certain nonlocal correlation function of the
underlying six-vertex model with domain wall boundary conditions, the
so-called emptiness formation probability (EFP).  We consider EFP in two
different representations: as a tau-function for Toda chains and as a random
matrix model integral.  The latter has a discrete measure and a linear
potential with hard walls; the observed phase transition shares properties
with both Gross-Witten-Wadia and Douglas-Kazakov phase transitions.
\end{abstract}

\maketitle
\section{Introduction}

Dimer coverings and random tilings of regular lattices are problems of great
and long-lasting interest \cite{FT-61,F-61,K-61,K-63,LW-99}. In spite of the
simplicity of their formulation, they exhibit numerous fascinating features. In
particular, the dependence of macroscopic quantities (such as the free energy)
on boundary conditions is most unusual and intriguing. For some recent advances
on physics of dimers and related tiling problems and their applications see,
e.g., \cite{GSBB-09,WTBG-12} and references therein.

The influence of boundary conditions on macroscopic quantities
is related with the fact that random tiling of finite planar regions
may exhibit phase separation phenomena. A famous example is provided by the domino
tilings of the ``Aztec diamond'', that exhibits ``frozen'' domino
configurations in the corners, outside a central region of disorder;
the phase separation curve emerging in the scaling limit is called
arctic circle, or arctic ellipse for a weighted counting of
configurations \cite{JPS-98} (for recent developments, see, e.g.,
\cite{CJY-12} and references therein).  Another famous example is
given by lozenges (rhombi) tilings of an hexagon, which are equivalent
to boxed plane partitions \cite{CLP-98}. In rather general settings
these and other similar problems in mathematical literature
are usually treated in terms of dimer
models on planar bipartite graphs \cite{KO-05,KOS-06,KO-06}.

The phase separation phenomena can also be observed in vertex models,
which can be treated as models of interacting dimers \cite{E-99}.
Most impressive results on this topic so far
have been provided by numerical experiments \cite{SZ-04,AR-05}.
Some exact analytical results for the six-vertex model were obtained in
\cite{CP-08,CP-09,CPZj-10}; to progress further on this problem
additional studies of random tilings seem to be very relevant.

In dealing with tiling problems one may wonder
how stable are the observed bulk properties of tilings,
such as free energy, phase
separation phenomena, etc, against various deformations of the
considered finite region.  While it is clear that preserving a peculiar
microscopic shape of the boundary of the region (e.g., the staircase
shape for the boundary of the Aztec diamond) is crucial for obtaining
the phase separation, one can also consider macroscopic
transformations of the shape of the region.

The purpose of the present paper is to address this problem
on the example of domino tilings. Specifically, we consider here
the domino tilings of the Aztec diamond with a cut-off corner of macroscopic square shape
and given size, and study the bulk properties of the tilings as the
size is varied.  We rely on the well-known correspondence between the
domino tilings of the Aztec diamond and the six-vertex model with
domain wall boundary conditions (DWBC) \cite{K-82,EKLP-92}. Here we
extend this correspondence to the domino tilings of the Aztec diamond
with a cut-off corner.  In this case, the six-vertex model is
considered with a square portion of the lattice removed, but again
with DWBC.

The partition function of the six-vertex model on this modified
lattice can be written as certain nonlocal correlation function of the
six-vertex model on the original lattice. This correlation function is
exactly what is known as the emptiness formation probability (EFP),
which can be viewed as a test function for total ferroelectric order
in a rectangular subregion in a corner of the original lattice.  Here
we specialize the subregion to a square, and derive the behavior of
EFP in the thermodynamic limit.

Restating the result in terms of tilings, we get the free energy of
the domino tilings of the Aztec diamond with a cut-off corner, as a
function of the size of the cut-off square. We observe that the free
energy exhibits a third-order phase transition when the cut-off
square, increasing in size, reaches the phase separation curve of the
original (unmodified) Aztec diamond, the arctic ellipse.

This result provides a novel insight on the phase separation phenomena
in tiling models. The phase separation curves can be seen as critical
curves in the space of parameters describing the macroscopic geometry
of the tiled region. While here we provide just a single example of
this interpretation, focusing on the domino tilings of Aztec diamond,
it may have rather universal nature and be observed in other tiling
problems and dimer models.

We organize the paper as follows. In the next section we discuss the
relation between tilings of the Aztec diamond with a cut-off square
and EFP in the six-vertex model.  In Section 3 we discuss a Hankel
determinant formula for EFP, which can be viewed both as a solution of
Toda chain differential equations, and as a random matrix model
integral (with a discrete measure).  In Section 4 we derive the result
using the differential equations approach. In Section 5 we show how
the same result can be obtained from the random matrix model integral.
In Section 6 we discuss the connection of the observed phase
transition with the arctic ellipse.

\section{Domino tilings and the six-vertex model}

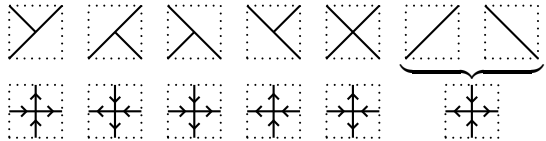
\begin{figure}
\centering
\psset{unit=10pt,dotsep=2pt}
\newcommand{\arr}{\lput{:U}{\begin{pspicture}(0,0)
\psline(-.05,.2)(.15,0)(-.05,-.2) \end{pspicture}}}


\begin{pspicture}(0,0)(20,5)
\rput(0,0){
\psline[linestyle=dotted](0,0)(0,2)(2,2)(2,0)(0,0)
\pcline(0,1)(1,1)\arr \pcline(1,1)(2,1)\arr
\pcline(1,0)(1,1)\arr \pcline(1,1)(1,2)\arr
}
\rput(3,0){
\psline[linestyle=dotted](0,0)(0,2)(2,2)(2,0)(0,0)
\pcline(1,1)(0,1)\arr \pcline(2,1)(1,1)\arr
\pcline(1,2)(1,1)\arr \pcline(1,1)(1,0)\arr
}
\rput(6,0){
\psline[linestyle=dotted](0,0)(0,2)(2,2)(2,0)(0,0)
\pcline(0,1)(1,1)\arr \pcline(1,1)(2,1)\arr
\pcline(1,2)(1,1)\arr \pcline(1,1)(1,0)\arr
}
\rput(9,0){
\psline[linestyle=dotted](0,0)(0,2)(2,2)(2,0)(0,0)
\pcline(2,1)(1,1)\arr \pcline(1,1)(0,1)\arr
\pcline(1,0)(1,1)\arr \pcline(1,1)(1,2)\arr
}
\rput(12,0){
\psline[linestyle=dotted](0,0)(0,2)(2,2)(2,0)(0,0)
\pcline(0,1)(1,1)\arr \pcline(2,1)(1,1)\arr
\pcline(1,1)(1,0)\arr \pcline(1,1)(1,2)\arr
}
\rput(16.5,0){
\psline[linestyle=dotted](0,0)(0,2)(2,2)(2,0)(0,0)
\pcline(1,1)(2,1)\arr \pcline(1,1)(0,1)\arr
\pcline(1,2)(1,1)\arr \pcline(1,0)(1,1)\arr
}
\rput(0,3){
\psline[linestyle=dotted](0,0)(0,2)(2,2)(2,0)(0,0)
\psline(0,0)(2,2)\psline(1,1)(0,2)
}
\rput(3,3){
\psline[linestyle=dotted](0,0)(0,2)(2,2)(2,0)(0,0)
\psline(0,0)(2,2)\psline(1,1)(2,0)
}
\rput(6,3){
\psline[linestyle=dotted](0,0)(0,2)(2,2)(2,0)(0,0)
\psline(0,2)(2,0)\psline(1,1)(0,0)
}
\rput(9,3){
\psline[linestyle=dotted](0,0)(0,2)(2,2)(2,0)(0,0)
\psline(0,2)(2,0)\psline(1,1)(2,2)
}
\rput(12,3){
\psline[linestyle=dotted](0,0)(0,2)(2,2)(2,0)(0,0)
\psline(0,0)(2,2)\psline(2,0)(0,2)
}
\rput(15,3){
\psline[linestyle=dotted](0,0)(0,2)(2,2)(2,0)(0,0)
\psline(0,0)(2,2)
}
\rput(18,3){
\psline[linestyle=dotted](0,0)(0,2)(2,2)(2,0)(0,0)
\psline(2,0)(0,2)
}
\rput(17.5,2.5){$\underbrace{\begin{pspicture}(-.2,0)(5.2,0)\end{pspicture}}$}
\end{pspicture}
\caption{Elementary patches of domino tilings (top) and the corresponding
configurations of the six-vertex model (bottom).}
\label{fig-DominoPatches}
\end{figure}

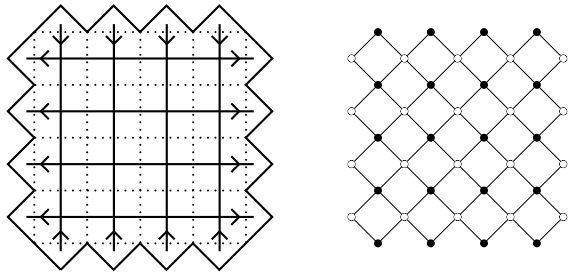
\begin{figure}
\centering


\psset{unit=20pt,dotsep=2pt}
\newcommand{\arr}{\lput{:U}{\begin{pspicture}(0,0)
\psline(-.1,.15)(.05,0)(-.1,-.15) \end{pspicture}}}

\begin{pspicture}(0,0)(10.5,5)
%
%
\rput(0,0){
\multirput(1,0)(1,0){4}{\pcline(0,.35)(0,1)\arr \pcline(0,4.65)(0,4)\arr}
\multirput(0,1)(0,1){4}{\pcline(1,0)(.35,0)\arr \pcline(4,0)(4.65,0)\arr}
\multirput(1,1)(0,1){4}{\psline(0,0)(3,0)}
\multirput(1,1)(1,0){4}{\psline(0,0)(0,3)}
\multirput(0,0)(1,0){5}{\psline[linestyle=dotted](.5,.5)(.5,4.5)}
\multirput(0,0)(0,1){5}{\psline[linestyle=dotted](.5,.5)(4.5,.5)}
\psline(1,0)(0,1)(.5,1.5)(0,2)(.5,2.5)(0,3)(.5,3.5)(0,4)(1,5)
  (1.5,4.5)(2,5)(2.5,4.5)(3,5)(3.5,4.5)(4,5)(5,4)
  (4.5,3.5)(5,3)(4.5,2.5)(5,2)(4.5,1.5)(5,1)(4,0)
  (3.5,.5)(3,0)(2.5,.5)(2,0)(1.5,.5)(1,0)
}

\rput(6,0){
\multirput(0,0)(1,0){4}{
\multirput(0,0)(0,1){4}{\psline[linewidth=.01](1,.5)(.5,1)(1,1.5)(1.5,1)(1,.5)}
}
\multirput(0,0)(0,1){5}{
\multirput(0,0)(1,0){4}{\psdot[dotsize=.15](1,.5)}}
\multirput(0,0)(1,0){5}{
\multirput(0,0)(0,1){4}{\psdot[dotsize=.15,dotstyle=o](.5,1)}}
}

\end{pspicture}
\caption{An Aztec diamond of order $N$ and the underlying $N\times N$
square lattice with DWBC, $N=4$ (left); the bipartite graph for the
related dimer model (right).}
\label{fig-AztecDiamond}
\end{figure}

First, recall (see \cite{EKLP-92}) that domino tilings can be formulated
in terms of the six-vertex model on a square lattice by mapping elementary
patches of domino tilings to arrow configurations as shown on
\figurename~\ref{fig-DominoPatches}. The Aztec diamond of order $N$ then
corresponds to the six-vertex model on an $N\times N$ lattice with a specific
choice of arrows on the boundary edges, known as the DWBC \cite{K-82}, see
\figurename~\ref{fig-AztecDiamond}, picture on the left. As a dimer model on a
bipartite graph, domino tilings of the Aztec diamond correspond to the
coverings of a square portion of a fishnet-like square lattice, see
\figurename~\ref{fig-AztecDiamond}, picture on the right. Obviously, the
results below thus apply not only to tilings but also to dimers.

Let $w_i$ denote the Boltzmann weight of the $i$th arrow configuration around a
vertex, $i=1,\dots,6$, as they appear on \figurename~\ref{fig-DominoPatches}
from left to right. The plain enumerations of domino tilings correspond to the
choice $w_1=\ldots=w_5=1$ and $w_6=2$. More generally, one may consider $w_i$'s
arbitrary but obeying the free-fermion condition,
\begin{equation}\label{FF}
w_1w_2+w_3w_4=w_5w_6.
\end{equation}
The partition function of the six-vertex model with DWBC on the $N\times
N$ lattice, $Z_N$, for weights satisfying the condition \eqref{FF}, has
an extremely concise form (see, e.g., \cite{EKLP-92}):
\begin{equation}
Z_N=w_5^{\frac{N(N-1)}{2}}w_6^{\frac{N(N+1)}{2}}.
\end{equation}
We shall use the following parameterization:
\begin{equation}\label{weights}
w_1=w_2=\sqrt{\rho(1-\alpha)},\qquad
w_3=w_4=\sqrt{\rho\alpha},\qquad
w_5=1,\qquad
w_6=\rho.
\end{equation}
Since in any configuration of the six-vertex model with DWBC the number of
vertices of type 6 is equal to the number of vertices of type 5 plus $N$, the
parameter $\rho$ is just an overall normalization of weights. In the
enumeration of domino tilings in terms of the six-vertex model configurations
$\rho=2$.  On the contrary, the parameter $\alpha$ is relevant. In the domino
tilings it describes the asymmetry between the two orientations, namely NE-SW
and NW-SE, of dominoes, giving them the weights $\sqrt{2(1-\alpha)}$ and
$\sqrt{2\alpha}$, respectively, in their weighted (``biased'') counting
\cite{JPS-98}.

\begin{figure}
\centering


\psset{unit=10pt,linewidth=.05}
\newcommand{\arr}{\lput{:U}{\begin{pspicture}(0,0)
\psline(0,.2)(.2,0)(0,-.2) \end{pspicture}}}

\begin{pspicture}(0,0)(11,11)

\multirput(4,0)(1,0){7}{\pcline(0,10.65)(0,10)\arr}
\multirput(1,0)(1,0){10}{\pcline(0,0.35)(0,1)\arr }
\multirput(0,1)(0,1){10}{\pcline(10,0)(10.65,0)\arr}
\multirput(0,1)(0,1){7}{\pcline(1,0)(.35,0)\arr}
\multirput(1,0)(1,0){3}{\pcline(0,7.65)(0,7)\arr}
\multirput(0,10)(0,-1){3}{\pcline(4,0)(3.35,0)\arr}

\multirput(4,8)(0,1){3}{\psline(0,0)(6,0)}
\multirput(1,1)(0,1){7}{\psline(0,0)(9,0)}

\multirput(1,1)(1,0){3}{\psline(0,0)(0,6)}
\multirput(4,1)(1,0){7}{\psline(0,0)(0,9)}
\psline(1,0)(0,1)
\multirput(1,0)(1,0){9}{\psline(0,0)(.5,.5)(1,0)}
\psline(10,0)(11,1)
\multirput(11,1)(0,1){9}{\psline(0,0)(-.5,.5)(0,1)}
\psline(11,10)(10,11)
\multirput(10,11)(-1,0){6}{\psline(0,0)(-.5,-.5)(-1,0)}
\psline(4,11)(3.5,10.5)
\psline(.5,7.5)(0,7)
\multirput(0,7)(0,-1){6}{\psline(0,0)(.5,-.5)(0,-1)}
\pspolygon[fillstyle=solid,fillcolor=lightgray]
(.5,7.5)(0,8)(.5,8.5)(0,9)(.5,9.5)(0,10)(1,11)(1.5,10.5)(2,11)
(2.5,10.5)(3,11)(3.5,10.5)(3,10)(3.5,9.5)(3,9)(3.5,8.5)(2.5,7.5)(2,8)(1.5,7.5)(1,8)

\psline[linewidth=.01](.5,9.5)(1.5,10.5)
\psline[linewidth=.01](.5,8.5)(2.5,10.5)
\psline[linewidth=.01](.5,7.5)(3.5,10.5)
\psline[linewidth=.01](1.5,7.5)(3.5,9.5)
\psline[linewidth=.01](1,10)(1.5,9.5)
\psline[linewidth=.01](1,9)(1.5,8.5)
\psline[linewidth=.01](2,10)(2.5,9.5)
\psline[linewidth=.01](2,9)(2.5,8.5)

\psline[linewidth=.075](3,8)(3.5,7.5)


\rput{-90}(-.5,9)
 {$\underbrace{\begin{pspicture}(0.2,0)(2.8,0)\end{pspicture}}$}
\rput(-1.25,9){$s$}
\rput{-90}(-.5,4)
 {$\underbrace{\begin{pspicture}(0.2,0)(6.8,0)\end{pspicture}}$}
\rput(-1.25,4){$r$}

\rput{-180}(7,11.5)
 {$\underbrace{\begin{pspicture}(0.2,0)(6.8,0)\end{pspicture}}$}
\rput(7,12.25){$r$}

\rput{-180}(2,11.5)
 {$\underbrace{\begin{pspicture}(0.2,0)(2.8,0)\end{pspicture}}$}
\rput(2,12.25){$s$}

\end{pspicture}
\caption{The Aztec diamond of order $r+s$, with a cut-off square of
size $s$, and the corresponding lattice for the six-vertex model,
$r=7$, $s=3$.}
\label{fig-CutoffCorner}
\end{figure}
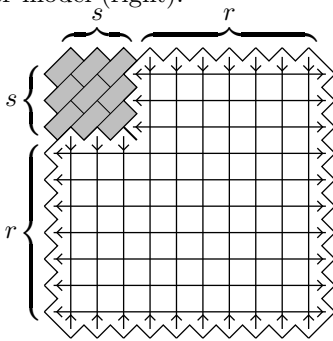

We now introduce the Aztec diamond with a cut-off corner.  Given an Aztec
diamond of order $N=r+s$, let us consider a new region which can be obtained
from the original Aztec diamond by removing the dashed subregion indicated in
\figurename~\ref{fig-CutoffCorner}.  This subregion can be viewed as an Aztec
diamond of order $s$, with one NE-SW row deleted in the middle, and it admits
only one tiling, with all dominoes NE-SW oriented.
The new region\footnote{Note that the new region must contain a line segment
attached to the SE corner of the removed subregion,
to keep connection with the six-vertex model.}
thus obtained is the Aztec diamond with a cut-off corner. We note that it can be
tiled with dominoes  only for $s\leq r$, as it can be easily seen in the dimer
context.

The Aztec diamond with a cut-off corner can be related to the six-vertex model
in standard way.  The six-vertex model is now defined on an $(r+s)\times(r+s)$
lattice with a square portion of the lattice, of size $s$, removed.  Due to the
relation between dominoes and vertex configurations
(\figurename~\ref{fig-DominoPatches}), the six-vertex model with a cut-off
square must have again DWBC, namely all horizontal external arrows are
outgoing, and all vertical ones are incoming, for the new boundary edges as
well, as shown in \figurename~\ref{fig-CutoffCorner}. We denote the partition
function of the six-vertex model on the new lattice as $Z_{r,s}$.

We are interested in the thermodynamic limit of large lattices, with large
sizes of the cut-off corner.  We thus consider both $r$ and $s$ large, with the
ratio $s/r=:v$ fixed.  Since the  model is meaningful only for $s\leq r$,
variable $v$ runs over the interval $[0,1]$.  We define the free energy per
site of the six-vertex model with a cut-off corner by
\begin{equation}\label{free}
F(v)=-\lim_{\substack{r,s\to\infty\\ s/r=v}}
\frac{\log Z_{r,s}}{r^2+2rs}.
\end{equation}
Clearly, $F(v)$ is also the free energy per domino for domino tilings of a
large Aztec diamond (of order $r(1+v)$, $r\to\infty$) with a cut-off corner of
square shape (and size $rv$).

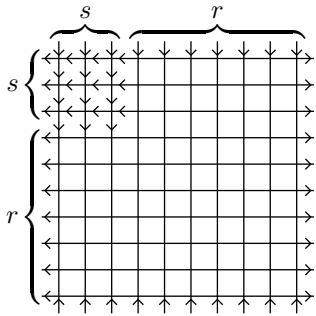
\begin{figure}
\centering
\psset{unit=10pt,linewidth=.05}
\newcommand{\arr}{\lput{:U}{\begin{pspicture}(0,0)
\psline(0,.2)(.2,0)(0,-.2) \end{pspicture}}}

\begin{pspicture}(-1,0)(11,12)
\multirput(1,0)(1,0){10}{\pcline(0,0.35)(0,1)\arr \pcline(0,10.65)(0,10)\arr}
\multirput(0,1)(0,1){10}{\pcline(1,0)(.35,0)\arr \pcline(10,0)(10.65,0)\arr}
\multirput(1,0)(1,0){3}{\pcline(0,10)(0,9)\arr}
\multirput(1,0)(1,0){3}{\pcline(0,9)(0,8)\arr}
\multirput(1,0)(1,0){3}{\pcline(0,8)(0,7)\arr}
 
\multirput(0,10)(0,-1){3}{\pcline(2,0)(1,0)\arr}
\multirput(0,10)(0,-1){3}{\pcline(3,0)(2,0)\arr}
\multirput(0,10)(0,-1){3}{\pcline(4,0)(3,0)\arr}

\multirput(4,8)(0,1){3}{\psline(0,0)(6,0)}
\multirput(1,1)(0,1){7}{\psline(0,0)(9,0)}

\multirput(1,1)(1,0){3}{\psline(0,0)(0,6)}
\multirput(4,1)(1,0){7}{\psline(0,0)(0,9)}

\rput{-90}(0,9)
  {$\underbrace{\begin{pspicture}(0.2,0)(2.8,0)\end{pspicture}}$}
\rput(-.75,9){$s$}
\rput{-90}(0,4)
  {$\underbrace{\begin{pspicture}(0.2,0)(6.8,0)\end{pspicture}}$}
\rput(-.75,4){$r$}

\rput{-180}(7,11)
  {$\underbrace{\begin{pspicture}(0.2,0)(6.8,0)\end{pspicture}}$}
\rput(7,11.75){$r$}

\rput{-180}(2,11)
  {$\underbrace{\begin{pspicture}(0.2,0)(2.8,0)\end{pspicture}}$}
\rput(2,11.75){$s$}

\end{pspicture}
\caption{The configuration of arrows whose probability is equal to $f_{r,s}$.}
\label{fig-EFP}
\end{figure}

To evaluate the partition function $Z_{r,s}$, let us now introduce EFP, a
nonlocal correlation function of the six-vertex model on the complete lattice
\cite{CP-07b}. This correlation function, denoted $f_{r,s}$, can be defined as
the probability that the vertices of the $s\times s$ subregion at the top-left
corner of the $(r+s)\times (r+s)$ lattice with DWBC are all of type 2, see
\figurename~\ref{fig-EFP}.  Note, that the definition of EFP can be easily
extended to a more general situation where the subregion at top-left corner has
a rectangular shape; here we consider only the case of a square shape.

It is clear that EFP coincides, modulo a simple overall factor, with the
partition function of the six-vertex model on the same $(r+s)\times (r+s)$
lattice but with all the vertices belonging to the $s\times s$ subregion in the
top-left corner lattice removed, see \figurename~\ref{fig-CutoffCorner}.
Indeed, for general values of the six-vertex model weights (i.e. not only under
the condition \eqref{FF}), we may write
\begin{equation}\label{Zrs}
Z_{r,s}=\frac{Z_{r+s}}{(w_2)^{s^2}}\; f_{r,s}.
\end{equation}
with the convention that $Z_{r,0}=Z_r$.

Given the above definition of EFP, one can compute it for some special cases.
First we note that the set of admissible configurations is empty for $s$
greater than $r$, so that EFP is nontrivial only for $s=0,1,\dots,r$. Aside
from the obvious case of $s=0$, in which
\begin{equation}\label{fr0}
f_{r,0}=1,
\end{equation}
there are two other easily computable cases, namely $s=1$ and $s=r$. In the
first case EFP can be inferred from the observation that the probability of
having a vertex of type 2 at the very top-left corner of the lattice is equal
to one, minus the probability of having the vertex of type 6 (see
\figurename~\ref{fig-DominoPatches}, and also discussion in \cite{BKZ-02}),
hence
\begin{equation}\label{fr1}
f_{r,1}=1-\frac{w_6 (w_3 w_4)^r Z_r}{Z_{r+1}}=1-\alpha^r.
\end{equation}
In the second case, $r=s$, the $2r\times 2r$ lattice splits onto four $r\times
r$ sublattices; all vertices of the top-left sublattice are of type 2, of the
bottom-right one are of type  1, and the top-right and the bottom-left
sublattices have DWBC, hence
\begin{equation}\label{frr}
f_{r,r}=\frac{(w_1w_2)^{r^2} Z_r^2}{Z_{2r}}=(1-\alpha)^{r^2}.
\end{equation}
Note that \eqref{fr0} and the first equalities in \eqref{fr1} and \eqref{frr}
hold for generic weights, independently of the free-fermion condition
\eqref{FF}.

In the thermodynamic limit, where both $r$ and $s$ are large, with the ratio
$v=s/r$ fixed, $v\in[0,1]$, EFP can be described by function $\sigma(v)$,
defined by
\begin{equation}\label{TDlimit}
f_{r,s}=\exp\left\{-r^2\sigma(s/r)+o(r^2)\right\},\qquad s,r\to\infty.
\end{equation}
Some properties of the function $\sigma(v)$ can be seen directly from the
definition of the EFP. For instance, $f_{r,s}$ being a probability, it varies
from $0$ to $1$, and thus $\sigma(v)\geq 0$. Furthermore, since increasing the
ratio $s/r=v$ corresponds to decreasing the number of configurations
contributing into the EFP, it follows that $\sigma(v)$ is a  nondecreasing
function of its variable.

From \eqref{weights} and \eqref{Zrs}, we may write the free energy per site
\eqref{free} of the six-vertex model with a cut-off corner as
\begin{equation}
F(v)=-\log\sqrt{\rho}+\frac{v^2}{1+2v}
\log\sqrt{1-\alpha}+\frac{1}{1+2v}\,\sigma(v).
\end{equation}
Thus the free energy density $F(v)$ is completely determined by the function
$\sigma(v)$, which is the object of our study in what follows.

Before addressing the exact form of the function $\sigma(v)$, let us
discuss its behaviour qualitatively.  We recall that the arctic circle
(or, in general, ellipse) phenomenon in the domino tilings of the Aztec diamond is the
emergence, in the thermodynamic limit, of four frozen regions in the
corners, sharply separated from a central region of disorder.

Turning to the language of the six-vertex model, this implies in
particular the presence of a frozen region of vertices of type 2 in
the top-left corner, outside the arctic ellipse.  Thus, by
construction (see also discussion in \cite{CP-07a}),
EFP must tend to one, as the lattice coordinates $r$ and $s$ get large,
$r,s\to\infty$, but such that the corresponding $s\times s$ subregion is entirely
contained in the top-left frozen region, i.e, outside of the arctic ellipse.
For the same reason, EFP is expected to tend to zero, as $r,s\to\infty$ and such
that the $s\times s$ subregion in the top-left corner overlaps with the central region
of disorder.

In view of this behavior, it is quite naturally to expect that
the function $\sigma(v)$, defined in
\eqref{TDlimit}, vanishes identically on some interval $[0,\vc]$,
where the value of $\vc$ corresponds to the arctic ellipse.
On the remaining interval, i.e., for $v\in [\vc,1]$, the function
$\sigma(v)$ is expected to be positive-valued and nondecreasing
function of $v$.  This is indeed what we shall observe in the following
by analysing some exact representation for EFP. Further discussion
in connection with the arctic ellipse phenomenon is given in Section 6.

\section{EFP as a Hankel determinant}

In this section we discuss our main tool to study EFP, namely, its
representation in terms of certain Hankel determinant.

We first mention that in our previous works, our treatment of EFP in the
six-vertex model with DWBC was based on certain representation in terms of a
multiple integral, valid for the model with generic weights, and derived in
\cite{CP-07b}.  In the special case of the free-fermion weights we used this
representation to recover the arctic ellipse in \cite{CP-07a}; the case of
generic weights and the corresponding arctic curves were considered in
\cite{CP-09,CPZj-10}.  However, such representation turns out having rather limited
capabilities to address the asymptotic properties of EFP in the thermodynamic
limit, even in the technically simple case of the free-fermion model (see,
e.g., the discussion in \cite{CP-07a}).

Motivated by this problem, in \cite{P-13} various alternative representations
for EFP in the case of the free-fermion model were provided. The main result
concerned a Hankel determinant formula, which can be viewed as resulting from
the evaluation of all the integrals in the multiple integral representation.
In the special case of EFP of a square shaped region, the Hankel determinant
representation reads:
\begin{equation}\label{frs}
f_{r,s}
=\frac{(1-\alpha)^{s^2}}{\big(\prod_{j=1}^{s-1} j!\big)^2\alpha^{s(s-1)/2}}
\,\det_{1\leq j,k\leq s} \left[\sum_{m=0}^{r-1}m^{j+k-2} \alpha^m\right].
\end{equation}
This representation allows one to relate EFP with a set of Toda chain
differential equations and with a random matrix model possessing a
simple potential.

A relation of \eqref{frs} with Toda chains can be established due to its Hankel
determinant structure, using the fact that all entries in the
determinant can be obtained by acting with the differential operators
$(\alpha\partial_\alpha)^{j+k-2}$ on certain function of $\alpha$,
independent of $s$.
This allows one to use the Sylvester determinant identity
to treat the determinant as the tau-function of some Toda chain
\cite{AP-87,KZj-00}. One can obtain the following equation for EFP:
\begin{equation}\label{TodaS}
\left(\alpha\partial_\alpha\right)^2
\log f_{r,s}
=\frac{s^2 \alpha}{(1-\alpha)^2}
\left(\frac{f_{r,s+1}f_{r,s-1}}{f_{r,s}^2}-1\right).
\end{equation}
This equation, supplemented by values of $f_{r,s}$ at $s=0$ and $s=1$, given by
\eqref{fr0} and \eqref{fr1}, respectively, allows one to reconstruct the
function $f_{r,s}$ iteratively for the remaining values of $s$.  In
\cite{P-13}, it was also proven that, for any given $s$, the function $f_{r,s}$
satisfies the following equation, where now $r$ varies:
\begin{equation}\label{TodaR}
\left(\alpha\partial_\alpha\right)^2
\log f_{r,s}
=\frac{r^2 \alpha}{(1-\alpha)^2}
\left(\frac{f_{r+1,s}f_{r-1,s}}{f_{r,s}^2}-1\right).
\end{equation}
The existence of this equation can be ascribed to the fact that EFP has one
more representation, similar to \eqref{frs}, but now in terms of an
$(r-s)\times (r-s)$ determinant with entries depending on $s$ (but not on $r$);
for further details, see \cite{P-13}, Section 3.

The Hankel determinant representation \eqref{frs} can be easily
related to random matrix models as well. Indeed, extracting the sums
from the determinant, which takes the Vandermonde form, and symmetrizing
the obtained expression, one gets
\begin{equation}\label{RMMrepr}
f_{r,s}
=\frac{(1-\alpha)^{s^2}}{s!\big(\prod_{j=1}^{s-1}j!\big)^2\alpha^{s(s-1)/2}}
\sum_{m_1=0}^{r-1}\dots\sum_{m_s=0}^{r-1}
\prod_{j<k}(m_k-m_j)^2  \prod_{j=1}^s\alpha^{m_j}.
\end{equation}
This expression can be viewed as the standard integral formula, although with a
discrete measure, for a random matrix model partition function with a linear
potential and two hard walls, see the discussion in Section 5. We mention that
a similar expression appeared also in certain random growth model \cite{J-00}.

To derive the function $\sigma(v)$ from the Hankel determinant representation,
we shall need its values, at $v=0$ and $v=1$, for the Toda differential
equation approach, and as $\alpha\to 0$ and $\alpha\to 1$, for the random
matrix model approach. Some of these values directly follows from the
definition of EFP, while other can be obtained only from an explicit evaluation
of the Hankel determinant in \eqref{frs}.

First, we consider EFP in the case of large $r$ and finite $s$. As explained in
Appendix A, the Hankel determinant in \eqref{frs} admits an explicit evaluation
at $r=\infty$, for $\alpha$ arbitrary.  The result is given by \eqref{meixner}.
Since for large $r$ every entry can be represented as its value at $r=\infty$,
minus an $O(\alpha^r)$ term, we have an estimate of the form
\begin{equation}
\det_{1\leq j <k\leq s}
\left[\sum_{m=0}^{r-1}m^{j+k-2}\alpha^m\right]
=\prod_{j=0}^{s-1}
\frac{(j!)^2\alpha^{j}}{(1-\alpha)^{2j+1}} + O(\alpha^r).
\end{equation}
In this estimate we assume that $\alpha\in(0,1)$, i.e., we  exclude the values
$\alpha=0$ and $\alpha=1$ in the discussion of the large $r$ limit, $s$ finite.
The formula above implies that EFP takes values close to one, up to
exponentially small corrections,
\begin{equation}\label{larger}
1-f_{r,s}=O(\alpha^r),\qquad r\to\infty.
\end{equation}
Such behavior in fact implies that at $v=0$ the function $\sigma(v)$ vanishes
together with all its derivatives,
\begin{equation}\label{sigmav0}
\sigma^{(n)}(0)=0, \qquad n=0,1,\dots.
\end{equation}
Note that for $n=0,1$ this could already be argued from \eqref{fr0} and
\eqref{fr1}.

Next we consider the case of $s=r$, at arbitrary $\alpha$. It can be easily
seen that \eqref{frs} immediately reproduces \eqref{frr} by noting that the
multiple sum in \eqref{RMMrepr}, modulo permutations, is given by a single
term, with $m_j=j-1$. Hence,
\begin{equation}\label{condensed}
\sum_{m_1=0}^{r-1}\dots\sum_{m_r=0}^{r-1}
\prod_{j<k}(m_k-m_j)^2  \prod_{j=1}^{r}\alpha^{m_j} =
r!\prod_{j=1}^{r-1} (j!)^2\cdot \alpha^{r(r-1)/2},
\end{equation}
and in the thermodynamic limit we have
\begin{equation}\label{sigmav1}
\sigma(1)=-\log(1-\alpha),
\end{equation}
which will be used together with \eqref{sigmav0} in the next section.

Let us now treat the case of $\alpha\to 0$, at arbitrary $r$ and $s$. The
result for EFP can be obtained directly from its definition, since the weights
$w_3$ and $w_4$ vanish, and there is just a single configuration contributing
to the partition function. We simply have
\begin{equation}\label{falpha0}
\lim_{\alpha\searrow 0}f_{r,s}=1.
\end{equation}
The same result is easily recovered from \eqref{RMMrepr} by noting that, for
small $\alpha$, only the term $m_j=j-1$ in the multiple sum (modulo
permutations) is relevant, since all other terms are of higher order in
$\alpha$, and \eqref{condensed} applies.  In the thermodynamic limit we have
\begin{equation}\label{sigmaalpha0}
\lim_{\alpha\searrow 0}\sigma(v)=0,
\end{equation}
that will be used in the random matrix model approach in Section 5.

Finally, we study the case of $\alpha\to 1$.  From the definition of EFP,
recalling that in this case the weights $w_1$ and $w_2$ vanish, and that in
each configuration these weights appear in pairs, it follows that
\begin{equation}\label{falpha1}
f_{r,s}\sim C_{r,s} (1-\alpha)^{s^2}, \qquad \alpha\to 1,
\end{equation}
where $C_{r,s}$ is some quantity independent of $\alpha$, which cannot be
worked out simply by inspecting the configurations.  However, one can find
$C_{r,s}$ from the Hankel determinant representation for EFP, since, as
explained in Appendix A, the determinant in \eqref{frs} at $\alpha=1$ can be
explicitly evaluated, with the result given by \eqref{alpha=1b}. Plugging it
into \eqref{frs} leads to \eqref{falpha1} where the quantity $C_{r,s}$ reads
\begin{equation}\label{crs}
C_{r,s}= \prod_{j=0}^{s-1}
\frac{(j!)^2 (j+r)!}{(2j)!(r-j-1)!(2j+1)!}.
\end{equation}
In the thermodynamic limit, the quantity of interest being $\sigma(v)$,
we may write
\begin{equation}\label{sigmaalpha1}
\lim_{\alpha\nearrow 1} \left[
\sigma(v)+v^2\log(1-\alpha)\right]=
\psi(v),
\end{equation}
where
\begin{equation}\label{defpsi}
\psi(v):=-\lim_{\substack{r,s\to\infty\\ s/r=v}}\frac{\log C_{r,s}}{r^2}.
\end{equation}
From the exact expression \eqref{crs} for the quantity $C_{r,s}$,
using standard arguments based on Stirling formula, we easily obtain
\begin{equation}\label{psi}
\psi(v)=v^2\log 4v
-\frac{(1-v)^2}{2}\log(1-v)-\frac{(1+v)^2}{2}\log(1+v).
\end{equation}
The function $\psi(v)$ enters the derivation of the function $\sigma(v)$
from the random matrix model integral.

\section{The differential equations approach}

The approach which we intend to apply here to derive the function $\sigma(v)$
is strongly inspired by that used in \cite{KZj-00} where the partition function
of the six-vertex model with DWBC was studied exploiting its property of being
the tau-function of a semi-infinite Toda chain.

The derivation is based on the idea that \eqref{TodaS} and
\eqref{TodaR} can be used to derive a set of partial differential
equations for $\sigma(v)$.
Indeed, substituting (the logarithm of) definition \eqref{TDlimit}
into \eqref{TodaS} and \eqref{TodaR}, replacing $s$ with $v r$,
dividing by $r^2$, and taking the limit $r\to\infty$, we obtain
\begin{equation}\label{PDEsigma}
\begin{split}
(\alpha\partial_\alpha)^2\sigma&=\frac{v^2 \alpha}{(1-\alpha)^2}
\left(1-\rme^{-\sigma''}\right),
\\
(\alpha\partial_\alpha)^2\sigma&=\frac{\alpha}{(1-\alpha)^2}
\left(1-\rme^{-v^2\sigma''+2v\sigma'-2\sigma}\right).
\end{split}
\end{equation}
Here the prime denotes the derivative with respect to $v$ and to
simplify writing we lift the dependence on $v$.
The limit performed above holds uniformly in $v$, over any interval
where $\sigma''$ exists.

We first obtain solutions of the system of equations \eqref{PDEsigma}, and next
explain how they can be used to construct the function $\sigma(v)$ describing
the thermodynamic limit of the EFP. Equating the right-hand sides of the two
equations in \eqref{PDEsigma} gives us the following ordinary differential
equation in $v$:
\begin{equation}\label{ODE}
v^2\rme^{(v^2-1)\sigma''-2v\sigma'+2\sigma}
+(1-v^2)\rme^{v^2\sigma''-2v\sigma'+2\sigma}=1.
\end{equation}
Since both terms here in the left-hand side are nonnegative and vary from $0$
to $1$, we can parameterize them as $\sin^2\!\varphi$ and $\cos^2\!\varphi$,
respectively, with function $\varphi=\varphi(v)$ taking values over the
interval $[0,\pi/2]$.  Thus we replace \eqref{ODE} by the system of two
equations for two functions, $\sigma$ and $\varphi$,
\begin{equation}\label{trigparam}
\begin{split}
\sigma''&=-2\log\frac{\sin\varphi}{v}+2\log\frac{\cos\varphi}{\sqrt{1-v^2}},
\\
v \sigma'-\sigma
&=-v^2\log\frac{\sin\varphi}{v}
-(1-v^2)\log\frac{\cos\varphi}{\sqrt{1-v^2}}.
\end{split}
\end{equation}
Since $(v\sigma'-\sigma)'=v\sigma''$, differentiating the second
equation and subtracting the first one multiplied by $v$ allows us to
eliminate $\sigma$ and to obtain the following equation for $\varphi$:
\begin{equation}
\left[(1-v^2)\tan \varphi-v^2\cot\varphi\right]\varphi'=0.
\end{equation}
Obviously, this equation has two solutions, which we denote as
$\varphi_\mathrm{I}$ and $\varphi_\mathrm{II}$, which read
\begin{equation}
\varphi_\mathrm{I}(v)=\arcsin v,\qquad
\varphi_\mathrm{II}(v)=\omega,
\end{equation}
where $\omega$ is some function of $\alpha$ (but not of $v$).

Denoting by $\sigma_\mathrm{I}$ the solution of
\eqref{trigparam} that corresponds to $\varphi_\mathrm{I}$, we find
from \eqref{trigparam} that it must satisfy the equations
$\sigma''_\mathrm{I}=v \sigma'_\mathrm{I}-\sigma_\mathrm{I}^{}=0$.
Hence, it is just a linear function in $v$ of the form
\begin{equation}\label{sigmaI}
\sigma_\mathrm{I}(v)=A v,
\end{equation}
where $A$ depends on $\alpha$. To fix $A$, we plug this solution of
\eqref{ODE} into \eqref{PDEsigma} that gives us the equation
$(\alpha\partial_\alpha)^2A=0$, which can be readily solved with the
result
\begin{equation}\label{coefA}
A=a_1+a_2\log\alpha,
\end{equation}
where $a_1$ and $a_2$ are arbitrary constants.

The solution $\sigma_\mathrm{II}$ of \eqref{trigparam} that
corresponds to $\varphi_\mathrm{II}$ can be worked out in a similar
manner. Solving \eqref{trigparam} with $\varphi$ replaced by a
$v$-independent function $\omega$, we find the expression
\begin{multline}\label{sigmaII}
\sigma_\mathrm{II}(v)
=v^2\log v-\frac{(1-v)^2}{2}\log(1-v)-\frac{(1+v)^2}{2}\log(1+v)
\\
+v^2\log\cot\omega+B v+\log\cos\omega,
\end{multline}
where $B$ depends on $\alpha$. We can find $B$ and $\omega$ by
substituting the obtained solution into one of the two equations in
\eqref{PDEsigma}. The resulting equation has in both sides second
order polynomials in $v$; matching terms in powers of $v$ we obtain
three ordinary differential equations in $\alpha$. The equation due
to the first order term in $v$ is $(\alpha\partial_\alpha)^2 B=0$,
from which we conclude that
\begin{equation}\label{bbb}
B=b_1+b_2 \log\alpha,
\end{equation}
where $b_1$ and $b_2$ are some constants. The remaining two equations,
due to the zeroth and second order terms in $v$, are
\begin{equation}
(\alpha\partial_\alpha)^2\log\cos\omega
=-\frac{\alpha}{(1-\alpha)^2}\tan^2\!\omega
\end{equation}
and
\begin{equation}
(\alpha\partial_\alpha)^2\log\cot\omega
=\frac{\alpha}{(1-\alpha)^2}\frac{1}{\cos^2\omega},
\end{equation}
respectively. Solving these equations and choosing the compatible
solution, which additionally satisfies the condition to be a real-valued
function of $\alpha$ taking values in the interval $[0,\pi/2]$, we
obtain that
\begin{equation}\label{omega}
\omega=\arcsin u,\qquad u:=\frac{1-\sqrt\alpha}{1+\sqrt\alpha}.
\end{equation}
This completes the construction of the solutions of the system of
equations \eqref{PDEsigma}.

Let us now establish the form of the function $\sigma(v)$ describing the
thermodynamic limit of the EFP.  First of all we note that the
function $\sigma(v)$ cannot be expressed in terms of either
$\sigma_\mathrm{I}(v)$ or $\sigma_\mathrm{II}(v)$ alone. Indeed,
while $\sigma_\mathrm{I}(v)$ can be easily made to satisfy the condition
at $v=0$, see \eqref{sigmav0}, it cannot satisfy the one at $v=1$, see
\eqref{sigmav1}, since the coefficient $A$, given by \eqref{coefA}, has
a different $\alpha$-dependence.  At the same time, function
$\sigma_\mathrm{II}(v)$ can be chosen to satisfy the boundary
condition at $v=1$, but it cannot be made consistent with the one at
$v=0$.

Thus, to satisfy the boundary conditions we have to assume
that the function $\sigma(v)$ is given by a combination of
the two solutions above, in agreement with the discussion at the end of Sect.~2,
as follows:
\begin{equation}\label{split}
\sigma(v)=
\begin{cases}
\sigma_\mathrm{I}(v) & v\in [0,\vc]
\\
\sigma_\mathrm{II}(v) & v\in [\vc,1].
\end{cases}
\end{equation}
Here $\vc$ is some function of the parameter $\alpha$.
As we show in the remaining part of this section, the assumption of the
structure of the function $\sigma(v)$ given by \eqref{split}
allows one to construct the unique solution
of \eqref{PDEsigma}, which possesses the required properties of being a
continuous, nonnegative and nondecreasing function of $v$
over the whole interval $[0,1]$, and satisfies
the boundary conditions \eqref{sigmav0} and \eqref{sigmav1}.

To completely determine our function $\sigma(v)$ according to
\eqref{split}, and find $\vc$, we first satisfy the boundary
conditions by fixing the constants $a_1$, $a_2$, $b_1$, and $b_2$
entering \eqref{sigmaI} and \eqref{sigmaII}. At $v=0$, see
\eqref{sigmav0}, we have the condition $A=0$, that is $a_1=a_2=0$.
Thus, the solution $\sigma_\mathrm{I}(v)$ relevant to our function
$\sigma(v)$ is just
\begin{equation}\label{solregimeI}
\sigma_\mathrm{I}(v)= 0.
\end{equation}
Let us now consider the point $v=1$, see \eqref{sigmav1}. From
\eqref{sigmaII}, \eqref{bbb} and \eqref{omega} we get
\begin{equation}
\sigma_\mathrm{II}(1)=b_1+ b_2\log\alpha+\log\frac{\sqrt\alpha}{1-\alpha}
\end{equation}
and hence \eqref{sigmav1} can be satisfied by choosing $b_1=0$ and
$b_2=-1/2$.  The resulting function $\sigma_\mathrm{II}(v)$ can be
written as
\begin{equation}\label{solregimeII}
\sigma_\mathrm{II}(v)=
v^2 \log\frac{v}{u}
-\frac{(1-v)^2}{2}\log\frac{1-v}{1-u}-\frac{(1+v)^2}{2}\log\frac{1+v}{1+u},
\end{equation}
where $u$ is defined in \eqref{omega}.

The value $\vc$ can now be determined from the requirement that the
function $\sigma(v)$, given by \eqref{split}, \eqref{solregimeI} and
\eqref{solregimeII}, is continuous, nonnegative and nondecreasing.
Since $\sigma_\mathrm{I}(v)$ is just equal to zero, continuity implies
that $\sigma_\mathrm{II}(\vc)=0$.  From \eqref{solregimeII} one can
see that $\sigma_\mathrm{II}(v)$ is positive and monotonously
increasing, from $0$ at $v=u$ to its boundary value $-\log(1-\alpha)$
at $v=1$.  This allows one to determine the value of the constant
$\vc$ in \eqref{split} to be
\begin{equation}\label{vcvalue}
v_c=u=\frac{1-\sqrt\alpha}{1+\sqrt\alpha}.
\end{equation}
Thus, the function $\sigma(v)$ is given by \eqref{split}, \eqref{solregimeI},
\eqref{solregimeII} and \eqref{vcvalue}.

As a simple verification of our result here, one can easily check that
the limiting conditions at $\alpha=0$ and at $\alpha=1$, given by
\eqref{sigmaalpha0} and \eqref{sigmaalpha1}, respectively, are indeed
satisfied.

\section{The random matrix model approach}

Here we exploit the approach of \cite{Zj-00}.  We start from
representation \eqref{RMMrepr}, which we write separating explicitly
the matrix model integral
\begin{equation}
f_{r,s}
=\frac{(1-\alpha)^{s^2}}{\alpha^{s(s-1)/2}}I_{r,s},
\end{equation}
where we have defined
\begin{equation}\label{RMMrepr2}
I_{r,s}
=\frac{1}{s!\big(\prod_{j=1}^{s-1}j!\big)^2}
\sum_{m_1=0}^{r-1}\dots\sum_{m_s=0}^{r-1}
\prod_{j<k}(m_k-m_j)^2
\prod_{j=1}^s\alpha^{m_j}.
\end{equation}
In this formula one can easily recognize the discrete measure analogue of
an Hermitian $s\times s$ random matrix integral in terms of its eigenvalues.
Introducing the rescaled
coordinate $R=r/s$, the large $s$ limit of $I_{r,s}$ can be described
by the function $\Phi(R)$, defined by
\begin{equation}\label{defPhi}
I_{r,s}=\exp\left\{s^2\Phi(r/s)+o(s^2)\right\}, \qquad s\to\infty.
\end{equation}
The functions $\sigma(v)$, defined by \eqref{TDlimit},
and $\Phi(R)$ are related by
\begin{equation}\label{sigmaPhi}
\sigma(v)=-v^2\log\frac{1-\alpha}{\sqrt\alpha}-v^2\Phi(1/v),
\end{equation}
and variables $v$ and $R$ are related by $v=1/R$.

To derive the function $\Phi(R)$ we rewrite \eqref{RMMrepr2},
introducing the rescaled variables $\mu_j$ as follows:
\begin{equation}\label{rescaling}
m_k=s \mu_j,\qquad j=1,\dots,s.
\end{equation}
After rescaling, sums can be reinterpreted as Riemann sums, and in the
large $s$ limit replaced by integrals, so that, as $s\to\infty$,
\begin{equation}\label{Irs}
I_{r,s}\sim
\varkappa_s\int_0^{R}\cdots \int_0^{R}
\prod_{j<k}(\mu_k-\mu_j)^2
\exp\Bigg\{s\log\alpha\sum_{j=1}^s\mu_j\Bigg\}\,
\rmd^s\mu,
\end{equation}
where the normalization constant $\varkappa_s$ can be easily inferred from
\eqref{RMMrepr2}, but is unessential for what follows.

Now the usual random matrix saddle-point analysis can be applied,
provided that one imposes a suitable additional constraint keeping
track of the discreteness of the $m_j$'s \cite{DK-93}, see also
\cite{BK-00,Zj-00}.  In \eqref{RMMrepr2}, all $m_j$'s must be
distinct, otherwise the Vandermonde determinant vanishes, and
therefore $|m_k-m_j|\geq 1$, for all $j\ne k$.  Introducing the
density $\rho(\mu)$ of the rescaled variables $\mu_j$, satisfying the
normalization condition
\begin{equation}\label{rhoz}
\int\rho(\mu)\,\rmd\mu=1,
\end{equation}
the constraint simply reads:
\begin{equation}\label{leqone}
\rho(\mu)\leq 1.
\end{equation}
In general, when the eigenvalues are trapped in a well of the
potential, they accumulate with maximal density at the bottom of the
well. In the present situation of discrete eigenvalues, due to
\eqref{leqone}, saturated regions, i.e., where $\rho(\mu)=1$, may arise.

Another phenomenon in the present situation is related to the presence
of ``hard walls'', i.e., the fact that the rescaled eigenvalues are
restricted to the interval $\mu_j\in[0,R]$. If the eigenvalues were
continuous, this would imply, in the case of nonvanishing density near
an hard wall, an inverse square root singularity for the density
\cite{B-65,NW-91,TW-94b,DM-06,DM-08,CK-08}.  Hence, due to the constraint
\eqref{leqone} in the case of discrete eigenvalues, in the vicinity of
the hard wall the density can only vanish or saturate \cite{DS-00}.

In the random matrix model picture \eqref{Irs}
we have the hard wall potential well
$V(\mu)=+\infty$, for $\mu\not\in[0,R]$, and
\begin{equation}\label{Vmu}
V(\mu)=-\mu\log\alpha, \qquad \mu\in[0,R].
\end{equation}
Note that the potential is linear for $\mu\in[0,R]$,
with a positive slope, since $\alpha\in(0,1)$.

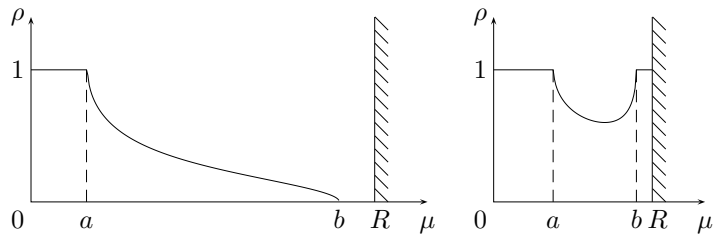
\begin{figure}
\centering
%

\psset{unit=10pt}
\begin{pspicture}(-1,-1)(26,7.5)
\psset{xunit=5,yunit=5,linewidth=.03}
\savedata{\uno}[
{{0., 1}, {0.01, 1}, {0.02, 1}, {0.03, 1}, {0.04, 1}, {0.05, 1}, {0.06, 1},
{0.07, 1}, {0.08, 1}, {0.09, 1}, {0.1, 1}, {0.11, 1}, {0.12, 1}, {0.13, 1},
{0.14, 1}, {0.15, 1}, {0.16, 1}, {0.17, 1}, {0.18, 1}, {0.19, 1}, {0.2, 1},
{0.21, 1}, {0.22, 1}, {0.23, 1}, {0.24, 1}, {0.25, 1}, {0.26, 1}, {0.27, 1},
{0.28, 1}, {0.29, 1}, {0.3, 1}, {0.31, 1}, {0.32, 1}, {0.33, 1}, {0.34, 1},
{0.35, 1}, {0.36, 1}, {0.37, 1}, {0.38, 1}, {0.39, 1}, {0.4, 1}, {0.41, 1},
{0.42, 1}, {0.43, 0.959359}, {0.44, 0.885831}, {0.45, 0.844706}, {0.46,
0.813154}, {0.47, 0.786846}, {0.48, 0.763994}, {0.49, 0.743648}, {0.5,
0.725231}, {0.51, 0.708358}, {0.52, 0.692759}, {0.53, 0.678233}, {0.54,
0.664628}, {0.55, 0.651825}, {0.56, 0.639726}, {0.57, 0.628253}, {0.58,
0.617341}, {0.59, 0.606934}, {0.6, 0.596987}, {0.61, 0.587457}, {0.62,
0.578311}, {0.63, 0.569518}, {0.64, 0.561051}, {0.65, 0.552887}, {0.66,
0.545003}, {0.67, 0.53738}, {0.68, 0.530003}, {0.69, 0.522855}, {0.7,
0.515922}, {0.71, 0.509192}, {0.72, 0.502653}, {0.73, 0.496294}, {0.74,
0.490105}, {0.75, 0.484078}, {0.76, 0.478204}, {0.77, 0.472475}, {0.78,
0.466885}, {0.79, 0.461426}, {0.8, 0.456093}, {0.81, 0.45088}, {0.82,
0.445781}, {0.83, 0.440791}, {0.84, 0.435906}, {0.85, 0.431121}, {0.86,
0.426432}, {0.87, 0.421835}, {0.88, 0.417326}, {0.89, 0.412902}, {0.9,
0.40856}, {0.91, 0.404296}, {0.92, 0.400108}, {0.93, 0.395992}, {0.94,
0.391946}, {0.95, 0.387968}, {0.96, 0.384055}, {0.97, 0.380205}, {0.98,
0.376415}, {0.99, 0.372684}, {1., 0.36901}, {1.01, 0.36539}, {1.02, 0.361824},
{1.03, 0.358308}, {1.04, 0.354842}, {1.05, 0.351424}, {1.06, 0.348052}, {1.07,
0.344726}, {1.08, 0.341442}, {1.09, 0.338201}, {1.1, 0.335002}, {1.11,
0.331841}, {1.12, 0.328719}, {1.13, 0.325635}, {1.14, 0.322587}, {1.15,
0.319574}, {1.16, 0.316595}, {1.17, 0.313649}, {1.18, 0.310736}, {1.19,
0.307854}, {1.2, 0.305002}, {1.21, 0.30218}, {1.22, 0.299386}, {1.23, 0.29662},
{1.24, 0.293882}, {1.25, 0.291169}, {1.26, 0.288483}, {1.27, 0.285821}, {1.28,
0.283183}, {1.29, 0.280569}, {1.3, 0.277977}, {1.31, 0.275408}, {1.32,
0.27286}, {1.33, 0.270333}, {1.34, 0.267827}, {1.35, 0.26534}, {1.36,
0.262872}, {1.37, 0.260423}, {1.38, 0.257993}, {1.39, 0.255579}, {1.4,
0.253183}, {1.41, 0.250804}, {1.42, 0.24844}, {1.43, 0.246092}, {1.44,
0.243759}, {1.45, 0.241441}, {1.46, 0.239137}, {1.47, 0.236847}, {1.48,
0.23457}, {1.49, 0.232305}, {1.5, 0.230053}, {1.51, 0.227814}, {1.52,
0.225585}, {1.53, 0.223368}, {1.54, 0.221162}, {1.55, 0.218966}, {1.56,
0.216779}, {1.57, 0.214603}, {1.58, 0.212435}, {1.59, 0.210276}, {1.6,
0.208126}, {1.61, 0.205983}, {1.62, 0.203848}, {1.63, 0.20172}, {1.64,
0.199599}, {1.65, 0.197485}, {1.66, 0.195376}, {1.67, 0.193273}, {1.68,
0.191175}, {1.69, 0.189082}, {1.7, 0.186993}, {1.71, 0.184909}, {1.72,
0.182828}, {1.73, 0.18075}, {1.74, 0.178675}, {1.75, 0.176602}, {1.76,
0.174531}, {1.77, 0.172462}, {1.78, 0.170393}, {1.79, 0.168326}, {1.8,
0.166258}, {1.81, 0.16419}, {1.82, 0.162121}, {1.83, 0.160051}, {1.84,
0.157979}, {1.85, 0.155904}, {1.86, 0.153827}, {1.87, 0.151746}, {1.88,
0.149661}, {1.89, 0.147571}, {1.9, 0.145476}, {1.91, 0.143375}, {1.92,
0.141267}, {1.93, 0.139152}, {1.94, 0.137028}, {1.95, 0.134895}, {1.96,
0.132753}, {1.97, 0.130599}, {1.98, 0.128434}, {1.99, 0.126256}, {2.,
0.124065}, {2.01, 0.121858}, {2.02, 0.119635}, {2.03, 0.117394}, {2.04,
0.115135}, {2.05, 0.112855}, {2.06, 0.110553}, {2.07, 0.108226}, {2.08,
0.105874}, {2.09, 0.103494}, {2.1, 0.101083}, {2.11, 0.0986384}, {2.12,
0.096158}, {2.13, 0.0936382}, {2.14, 0.0910752}, {2.15, 0.088465}, {2.16,
0.0858027}, {2.17, 0.0830831}, {2.18, 0.0802996}, {2.19, 0.0774452}, {2.2,
0.0745112}, {2.21, 0.0714872}, {2.22, 0.068361}, {2.23, 0.0651174}, {2.24,
0.0617372}, {2.25, 0.0581962}, {2.26, 0.0544625}, {2.27, 0.0504929}, {2.28,
0.0462259}, {2.29, 0.0415694}, {2.3, 0.0363733}, {2.31, 0.0303612}, {2.32,
0.0228976}, {2.33, 0.0114224}}
]

\rput(0,0){\dataplot{\uno}}

\psset{unit=1}
\psline{<->}(0,7)(0,0)(15,0)
\psline(13,0)(13,7)
\rput[rB](-.25,-1){$0$}
\rput[B](13.2,-1){$R$}
\rput[B](2.1,-1){$a$}
\rput[B](11.66,-1){$b$}
\rput[r](-.25,5){$1$}
\rput[B](15,-1){$\mu$}
\rput[r](-.25,7){$\rho$}
\psline[linestyle=dashed,linewidth=.01](2.1,5)(2.1,0)
\multirput(13,0)(0,0.5){14}{\psline(.5,0)(0,.5)}

\psset{xunit=5,yunit=5,linewidth=.03}

\savedata{\due}[
{{0., 1}, {0.01, 1}, {0.02, 1}, {0.03, 1}, {0.04, 1}, {0.05,
  1}, {0.06, 1}, {0.07, 1}, {0.08, 1}, {0.09, 1}, {0.1, 1}, {0.11,
  1}, {0.12, 1}, {0.13, 1}, {0.14, 1}, {0.15, 1}, {0.16, 1}, {0.17,
  1}, {0.18, 1}, {0.19, 1}, {0.2, 1}, {0.21, 1}, {0.22, 1}, {0.23,
  1}, {0.24, 1}, {0.25, 1}, {0.26, 1}, {0.27, 1}, {0.28, 1}, {0.29,
  1}, {0.3, 1}, {0.31, 1}, {0.32, 1}, {0.33, 1}, {0.34, 1}, {0.35,
  1}, {0.36, 1}, {0.37, 1}, {0.38, 1}, {0.39, 1}, {0.4, 1}, {0.41,
  1}, {0.42, 1}, {0.43, 1}, {0.44, 1}, {0.45, 1}, {0.46,
  0.929139}, {0.47, 0.885458}, {0.48, 0.855297}, {0.49,
  0.831213}, {0.5, 0.810846}, {0.51, 0.793073}, {0.52,
  0.777255}, {0.53, 0.762985}, {0.54, 0.749984}, {0.55,
  0.738051}, {0.56, 0.727035}, {0.57, 0.716822}, {0.58,
  0.707319}, {0.59, 0.698452}, {0.6, 0.69016}, {0.61,
  0.682395}, {0.62, 0.675114}, {0.63, 0.668281}, {0.64,
  0.661867}, {0.65, 0.655847}, {0.66, 0.650198}, {0.67,
  0.644903}, {0.68, 0.639945}, {0.69, 0.635311}, {0.7,
  0.63099}, {0.71, 0.626973}, {0.72, 0.623252}, {0.73,
  0.619822}, {0.74, 0.616678}, {0.75, 0.613818}, {0.76,
  0.611241}, {0.77, 0.608947}, {0.78, 0.606938}, {0.79,
  0.605216}, {0.8, 0.603788}, {0.81, 0.602659}, {0.82,
  0.601839}, {0.83, 0.601337}, {0.84, 0.601166}, {0.85,
  0.601342}, {0.86, 0.601883}, {0.87, 0.602809}, {0.88,
  0.604146}, {0.89, 0.605923}, {0.9, 0.608173}, {0.91,
  0.610939}, {0.92, 0.614267}, {0.93, 0.618214}, {0.94,
  0.622849}, {0.95, 0.628253}, {0.96, 0.634526}, {0.97,
  0.64179}, {0.98, 0.650198}, {0.99, 0.659945}, {1., 0.671281}, {1.01,
   0.684539}, {1.02, 0.700177}, {1.03, 0.718851}, {1.04,
  0.741555}, {1.05, 0.769947}, {1.06, 0.807199}, {1.07,
  0.861423}, {1.08, 1}, {1.09, 1}, {1.1, 1}, {1.11, 1}, {1.12,
  1}, {1.13, 1}, {1.14, 1}, {1.15, 1}, {1.16, 1}, {1.17, 1}, {1.18,
  1}, {1.19, 1}, {1.2, 1}}
]

\rput(3.5,0){
\dataplot{\due}}

\psset{unit=1}
\rput(17.5,0){
\psline{<->}(0,7)(0,0)(8,0)
\psline(6,0)(6,7)
\rput[rB](-.25,-1){$0$}
\rput[B](6.2,-1){$R$}
\rput[B](2.25,-1){$a$}
\rput[B](5.4,-1){$b$}
\rput[r](-.25,5){$1$}
\rput[B](8,-1){$\mu$}
\rput[r](-.25,7){$\rho$}
\psline[linestyle=dashed,linewidth=.01](2.25,5)(2.25,0)
\psline[linestyle=dashed,linewidth=.01](5.4,5)(5.4,0)
\multirput(6,0)(0,0.5){14}{\psline(.5,0)(0,.5)}
}
\end{pspicture}
\caption{Plots of densities in the two scenarios.}
\label{fig-DensityPlots}
\end{figure}

In view of the previous discussion, it is clear that according to the values of
the parameters $\alpha$ and $R$ two different scenarios can manifest. Let us
start with considering very large values of $R$ (and some generic fixed value
of $\alpha\not=0,1$). Since the eigenvalues accumulate to the bottom of the
linear potential, a saturated region $\rho(\mu)=1$ arises near the origin. At
the same time, the wall at $\mu=R$ is very far on the right and can be ignored.
Introducing positive real parameters $a$ and $b$, with $0<a<b<R$, to be determined
later, we thus have a saturated region $\rho(\mu)=1$ on some interval
$\mu\in[0,a]$, with an unsaturated region $\mu\in[a,b]$, and vanishing
eigenvalue density for $\mu\in[b,R]$, see \figurename~\ref{fig-DensityPlots},
picture on the left.

As $R$ decreases, as far as it remains larger than $b$, nothing
changes in the previous scenario. But as soon as $b=R$, the eigenvalue
density is constrained to jump from $\rho(R)=0$ to $\rho(R)=1$ and a
new scenario arises, with a saturated region on some interval
$\mu\in[0,a]$, an unsaturated region $\mu\in[a,b]$, and a second
saturated region for $\mu\in[b,R]$, see
\figurename~\ref{fig-DensityPlots}, picture on the right.

As $R$ decreases further and approaches the value $R=1$, the number of
available positions becomes smaller and smaller, and eventually barely
sufficient to accommodate the $s$ distinct eigenvalues.
Correspondingly, the central unsaturated region of the second scenario
shrinks down. In particular, at $R=1$, the central unsaturated region
disappears completely, and $\rho(\mu)=1$, for $\mu\in[0,1]$. At the
end, there is no admissible eigenvalue configuration for $R<1$.

To proceed, we recall that in the saddle-point approximation the eigenvalue
density is related to the resolvent
\begin{equation}\label{Wdef}
W(z)=\int_{S}\frac{\rho(\mu)}{z-\mu}\,\rmd\mu,\qquad z\not\in S,
\end{equation}
in particular, to its discontinuity across its cut $S$,
\begin{equation}\label{rhodef}
\rho(z)=-\frac{1}{2\pi\rmi}\left[W(z+\rmi 0)-W(z-\rmi 0)\right], \qquad z\in S.
\end{equation}
Clearly, $S$ is also the support of the density $\rho(\mu)$.
In turn, the resolvent is  determined  by  the saddle-point equation
\begin{equation}\label{spegen}
W(z+\rmi 0)+W(z-\rmi 0)=U(z), \qquad    z\in S.
\end{equation}
In the case of a continuous measure, $U(z)$ is just equal to the
derivative $V'(z)$, where $V(z)$ is the potential of the model.
However, in the case of discrete measure, this holds only as far as
the density $\rho(z)$ does not saturate the constraint \eqref{leqone}.
The occurrence of saturation in the density requires modifying the form
of $U(z)$ and $S$ in \eqref{spegen}, as we shall discuss later on.

Assuming that $S$ consists of a single interval $[a,b]$ on the real
axis, the solution of \eqref{spegen} that is consistent with \eqref{Wdef}
is
\begin{equation}\label{Wint}
W(z)=\frac{\sqrt{(z-a)(z-b)}}{2\pi}
\int_a^b\frac{U(u)}{(z-u)\sqrt{(u-a)(b-u)}}\,\rmd u,
\end{equation}
Imposing the large $z$ asymptotic
behavior implied by \eqref{Wdef} together with \eqref{rhoz}, on the
order $z^0$ and $z^{-1}$ terms of the large $z$ expansion of
\eqref{Wint} fixes the endpoints $a$ and $b$ of the interval $S$.

Denoting by $E$ the average of the eigenvalues (i.e., the first moment
of the density $\rho(z)$), we note that it can be extracted from the
order $z^{-2}$ coefficient,
\begin{equation}\label{largez}
W(z)=\frac{1}{z}+\frac{E}{z^2}+ O(z^{-3}),\qquad |z|\to\infty.
\end{equation}
In the case of potential  of the form
\eqref{Vmu}, the average $E$ can be related to the function $\Phi(R)$,
defined in \eqref{defPhi}, by
\begin{equation}\label{E}
\alpha\partial_\alpha \Phi(R)=E,
\end{equation}
allowing one to determine $\Phi(R)$ up to some quantity independent
of $\alpha$.

Now we find solutions $W_\mathrm{I}(z)$ and $W_\mathrm{II}(z)$ of the
saddle-point equations corresponding to the two scenarios outlined
above, respectively. We also compute the corresponding functions
$\Phi_\mathrm{I}(z)$ and $\Phi_\mathrm{II}(z)$, and show that they
indeed reproduce \eqref{solregimeI} and \eqref{solregimeII}, as
expected.

Let us consider the first scenario, with one saturated region, i.e.,
with $\rho(\mu)=1$, for $\mu\in[0,a]$. The saddle-point problem in
this case coincides with that arising in the matrix model associated
to the partition function of the domain wall six-vertex model in
its ferroelectric phase, see \cite{Zj-00}, Section 3 (see also
\cite{J-00,BK-00,BL-07}).  The saturated region in the interval
$[0,a]$ gives rise to a logarithmic cut in the resolvent
$W_\mathrm{I}(z)$, which can be removed by introducing an auxiliary
function $H_\mathrm{I}(z)$ as follows:
\begin{equation}\label{WHI}
W_\mathrm{I}(z)=\log\frac{z}{z-a}+H_\mathrm{I}(z).
\end{equation}
The saddle-point equation for $H_\mathrm{I}(z)$ reads
\begin{equation}\label{spe1}
H_\mathrm{I}(z+\rmi0)+H_\mathrm{I}(z-\rmi0)
=-2\log\frac{\sqrt\alpha z}{z-a},\qquad z\in[a,b].
\end{equation}
Exploiting \eqref{spegen} and \eqref{Wint}, and evaluating
the resulting integral (see Appendix B), one has:
\begin{equation}\label{WI}
W_\mathrm{I}(z)=-\log\sqrt\alpha
-2\log\frac{\sqrt{a(z-b)}+\sqrt{b(z-a)}}{\sqrt{(b-a)z}}.
\end{equation}
Imposing the asymptotic behavior \eqref{largez} provides the
conditions
\begin{equation}
\frac{\sqrt{b}-\sqrt{a}}{\sqrt{b}+\sqrt{a}}=\sqrt{\alpha},\qquad
\sqrt{ab}=1,
\end{equation}
with the solution
\begin{equation}
a=\frac{1-\sqrt{\alpha}}{1+\sqrt{\alpha}},\qquad
b=\frac{1+\sqrt{\alpha}}{1-\sqrt{\alpha}}.
\end{equation}
Recall that the scenario holds as long as $R> \Rc$. The critical value
$\Rc$, at which the transition to the the second scenario
takes place, corresponds to $\Rc=b$, where the value for $b$ is
given above. Hence,
\begin{equation}\label{Rc}
\Rc=\frac{1+\sqrt{\alpha}}{1-\sqrt{\alpha}}.
\end{equation}
Recalling that $v=1/R$, this obviously reproduces \eqref{vcvalue}.

To conclude the discussion of the first scenario, we evaluate the
function $\Phi_\mathrm{I}(R)$. Expanding \eqref{WI} up to the order
$z^{-2}$, we get
\begin{equation}
E_\mathrm{I}=\frac{a+b}{4}=\frac{1+\alpha}{2(1-\alpha)}.
\end{equation}
Integration yields
\begin{equation}
\Phi_\mathrm{I}(R)=\log\frac{\sqrt{\alpha}}{1-\alpha},
\end{equation}
where the integration constant is fixed to comply with the condition
\eqref{sigmaalpha0}. As a result, see \eqref{sigmaPhi},
we reproduce \eqref{solregimeI}.

Let us now turn to the second scenario, with two saturated regions,
i.e., with $\rho(\mu)=1$, for $\mu\in[0,a] \cup [b,R]$. We first
remove the logarithmic cuts of $W_\mathrm{II}(z)$ arising from the
saturated regions,
\begin{equation}\label{WHII}
W_\mathrm{II}(z)=\log\frac{z(z-b)}{(z-a)(z-R)}+H_\mathrm{II}(z).
\end{equation}
The saddle-point equation for $H_\mathrm{II}(z)$ reads
\begin{equation}\label{spe2}
H_\mathrm{II}(z+\rmi 0)+H_\mathrm{II}(z-\rmi 0)
=-2\log\frac{\sqrt{\alpha}z(z-b)}{(z-a)(z-R)},
\qquad z\in[a,b].
\end{equation}
Using \eqref{Wint}, and evaluating the resulting integral (see
Appendix B), we obtain
\begin{equation}\label{WII}
W_\mathrm{II}(z)=-\log\sqrt\alpha
-\log\frac{z-R}{z}
-2\log\frac{\sqrt{a(z-b)}+\sqrt{b(z-a)}}
{\sqrt{R-a}\sqrt{z-b}+\sqrt{R-b}\sqrt{z-a}}.
\end{equation}
Imposing the asymptotic behavior \eqref{largez}
we obtain the two conditions,
\begin{equation}
\begin{split}
\bigg(\frac{\sqrt{R-a}+\sqrt{R-b}}{\sqrt{b}+\sqrt{a}}\bigg)^2
&=\sqrt{\alpha},
\\
\sqrt{ab}+\sqrt{(R-a)(R-b)}
&=1,
\end{split}
\end{equation}
with the solution
\begin{equation}\label{abII}
a=\frac{\big(\sqrt{R+1}-\sqrt{(R-1)\sqrt{\alpha}}\big)^2}{2(1+\sqrt{\alpha})},
\qquad
b=\frac{\big(\sqrt{R+1}+\sqrt{(R-1)\sqrt{\alpha}}\big)^2}{2(1+\sqrt{\alpha})}.
\end{equation}
The critical value $\Rc$, which corresponds to the case $b=R$,
reproduces \eqref{Rc}, as it should.

Now we are ready to compute the function $\Phi_\mathrm{II}(R)$.
Calculation of the $z^{-2}$ order term in the large $z$ expansion of
\eqref{WII} provides the first moment of the eigenvalue density:
\begin{equation}
E_\mathrm{II}=
\frac{a+b}{4}+\frac{R}{2}\sqrt{(R-a)(R-b)}.
\end{equation}
Substituting here $a$ and $b$ from \eqref{abII}, and integrating
\eqref{E}, we obtain the expression
\begin{equation}\label{PhiII}
\Phi_\mathrm{II}(R)=
(R^2-1)\log\frac{1+\sqrt\alpha}{2\alpha^{1/4}}
+R\log\sqrt\alpha+C(R).
\end{equation}
Here the integration constant $C(R)$, a quantity independent of
$\alpha$, can be determined due to the condition \eqref{sigmaalpha1},
with the result $C(R)=-R^2\psi(1/R)$. Recalling \eqref{sigmaPhi}, it
is immediately seen that \eqref{PhiII} indeed reproduces the result of
the previous section, \eqref{solregimeII}, for the function
$\sigma(v)$.

\section{Discussion}

Let us first summarize the results.  Our main result concerns the
explicit form of the function $\sigma(v)$, describing the
thermodynamic limit behavior of EFP. In deriving this result we used
two different methods. The first method is based on the connection of
the Hankel determinant representation for EFP with Toda chain
differential equations, and allows one to derive the result by rather
elementary means. The second method exploits the fact that the Hankel
determinant can be represented as a random matrix model integral. An
unusual feature of the matrix model arising in our study is the presence
of a discrete measure on a finite interval.

For the function $\sigma(v)$ we have derived the following expression:
\begin{equation}\label{sigma}
\sigma(v)=
\begin{cases}
0 &\quad  v\in[0,\vc]
\\
v^2 \log\dfrac{v}{\vc}
-\dfrac{(1-v)^2}{2}\log\dfrac{1-v}{1-\vc}
-\dfrac{(1+v)^2}{2}\log\dfrac{1+v}{1+\vc}
&\quad v\in[\vc,1].
\end{cases}
\end{equation}
The critical value $\vc=\vc(\alpha)$, at which function $\sigma(v)$
changes its behavior is
\begin{equation}\label{vcalpha}
\vc=\frac{1-\sqrt{\alpha}}{1+\sqrt{\alpha}}.
\end{equation}
The formulas above show that $\sigma'(v)$ and $\sigma''(v)$, where the
prime denotes derivative, are continuous functions in the vicinity of
$v=\vc$, with the values $\sigma'(\vc)=\sigma''(\vc)=0$, while the
third derivative has a discontinuity, since
\begin{equation}
\lim_{v\searrow\vc}\sigma'''(v)=\frac{1}{\vc(1-\vc^2)}.
\end{equation}
Thus at $v=\vc$ the function $F(v)$, which is the free energy per
domino for the domino tilings of the Aztec diamond with a cut-off
corner, has a discontinuity in its third derivative, i.e., at $v=\vc$
the model undergoes a third-order phase transition with respect to the
scaled size of the cut-off corner.

\begin{figure}
\centering

\psset{unit=10pt}
\begin{pspicture}(-3,-2.5)(12.5,11.5)
\psline(10,10)(10,0)(0,0)
\psline{->}(0,10)(12,10)
\psline{->}(0,10)(0,-2)
\rput[r](-.5,-2){$y$}
\rput[B](12,10.5){$x$}
\rput{45}(5,5){\psellipse[linewidth=.05](0,0)(5.5,4.4)}
\rput[rB](-.5,10.5){$0$}
\rput[B](10,10.5){$1$}
\rput[r](-.5,0){$1$}
\rput[r](-.5,3.9){$1{-}\alpha$}\psline[linewidth=.05](-.1,3.9)(.1,3.9)

\psline[linewidth=.15](0,8.1)(1.9,8.1)(1.9,10)
\psline[linewidth=.15](0,8.1)(0,0)(10,0)(10,10)(1.9,10)
\rput[r](-.5,8.1){$\frac{1{-}\sqrt\alpha}{2}$}

\end{pspicture}

\caption{A meaning of $\vc$: the cut off corner hits
the arctic ellipse of the initial Aztec diamond,
$\vc/(1+\vc)=(1-\sqrt\alpha)/2$.}
\label{fig-ArcticEllipse}
\end{figure}
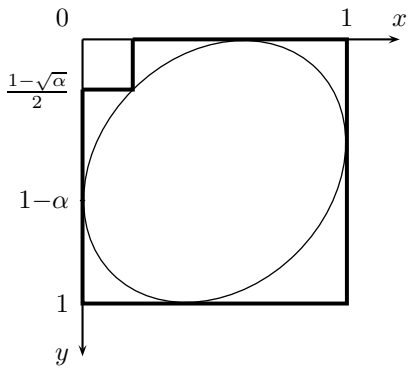

Let us now discuss the results. We first point out the meaning of the
change of the behavior of the function $\sigma(v)$ at $v=\vc$. The
value $v=\vc$ has a simple interpretation in terms of the original
model, i.e.  the unmodified  Aztec diamond, where frozen and temperate
regions are separated by the so-called arctic ellipse,
\begin{equation}\label{ellipse}
\frac{(1-x-y)^2}{\alpha}+\frac{(x-y)^2}{1-\alpha}=1.
\end{equation}
The value of $\vc$ given by \eqref{vcalpha} exactly corresponds to the
situation in which the cut-off corner, increasing in size, gets large
enough to reach the arctic ellipse of the original (unmodified) Aztec
diamond, see \figurename~\ref{fig-ArcticEllipse}.

This result suggests to view the phase separation curves as critical
curves in the space of parameters describing the macroscopic geometry
of the tiled region.  Indeed, while here we have restricted ourselves
to the case of a square shaped cut-off corner, one could consider the
more general situation of a rectangular shaped cut-off corner. The
treatment is essentially the same, with the free energy now exhibiting
a whole critical line, which coincides with the full portion of the
ellipse \eqref{ellipse} between its contact points with the top and
left boundaries. While finding the critical curve is straightforward,
the evaluation of the free energy is more involved, and will be
reported elsewhere.

Coming back to the discussion of a third-order phase transition in
random tilings, \eqref{sigma} can also be interpreted in a slightly
different manner, treating the size of the cut-off corner as some
given parameter, fixed from the beginning. Then, changing the
Boltzmann weights, that is varying the parameter $\alpha$, we again
have a third-order phase transition, as it can be easily verified from
\eqref{sigma}, when the ellipse, deforming itself as $\alpha$
is increasing, reaches the bottom-right vertex of the cut-off corner.

Recalling that the parameter $\alpha$ tunes the asymmetry between the
two possible orientations of dominoes, we see that the last
interpretation regards the arctic curves as critical curves in the
space of external fields acting on the system. A similar point of view
has already been considered, e.g., in the six-vertex model, where the
limit shape of the system has equivalently been studied in the space
parameterized by some external electric fields acting on it
\cite{RP-06}.

Interestingly enough, besides random tilings, this interpretation may
also have implications in the context of quantum spin chains, in view
of the well-known relation between the six-vertex model and the 1D
quantum Heisenberg spin chain.  Remnants of the arctic phenomena can
be seen when EFP of the quantum spin chain is treated by conformal
field theory methods \cite{S-13}. A possible third-order phase
transition in the quantum spin chain, although for a very peculiar
choice of the macroscopic parameters, has been pointed out recently in
\cite{PgT-13}. Third-order phase transitions have also been observed
in the related context of vicious walkers \cite{SMCF-13}.

Our last comment here concerns the observed phase transition
from the point of view of the random matrix model description.
In this respect we recall two
other third-order phase transitions related to random matrix models,
namely the Gross-Witten-Wadia \cite{GW-80,W-80}, and the
Douglas-Kazakov  \cite{DK-93} phase transitions, occurring in the large
$N$ limit of $U(N)$ Yang-Mills theory in two dimensions, when
formulated on the lattice, or in the continuum, respectively.

From the point of view of the random matrix model picture, in the
Gross-Witten-Wadia transition, the eigenvalues live on the unit circle, and
the two phases correspond to the support of the eigenvalue density
extending to the whole circle, or being restricted to an arc.  In the
Douglas-Kazakov transition the eigenvalues are discrete, thus imposing an
upper limit on the eigenvalue density. The two phases correspond to
the presence or absence of saturation in the eigenvalue density.

The matrix model phase transition observed here in relation to random
tilings shares properties with both Gross-Witten-Wadia and
Douglas-Kazakov phase transitions. Indeed, the transition can be
attributed to the rise of a saturated region, just like in the
Douglas-Kazakov transition.  At the same time, the two phases
correspond to the support of the eigenvalue density extending to the
whole allowed interval for the eigenvalues or restricting to a subset
of it, like in the Gross-Witten-Wadia transition.  We note further
that a common feature of these transitions is that they all
correspond to the appearance or disappearance of some edge of the
support of the (unsaturated part of the) eigenvalue density. The order
of the transition can be ascribed to the square-root behaviour of the
density at these edges.

In conclusion, we have discussed a third-order phase transition
arising in the random tilings of the Aztec diamond with a cut-off
corner, by deriving the leading term of asymptotics of a particular
correlation function in the closely related case of the six-vertex
model.  While here we have considered only the domino tilings of the
Aztec diamond with a cut-off corner, we believe that the phenomenon is
rather universal, and may be observed similarly in other tiling
problems where the phase separation phenomena are known to take place,
e.g., in the rhombus tilings of an hexagon with a cut-off rhombus, and
many others.

\section*{Acknowledgments}

We are indebted to A. Kuijlaars and N. Reshetikhin for useful
discussions.  This work is partially supported by the IRSES grant of
EC-FP7 Marie Curie Action ``Quantum Integrability, Conformal Field
Theory and Topological Quantum Computation'' (QIFCT).  A.G.P.
acknowledges partial support from the Russian Foundation for Basic
Research (grant 13-01-00336) and from INFN, Sezione di Firenze.

\appendix
\section{Evaluation of determinants}

Here we outline the  evaluation of two determinants used in Section 3.
Namely, the determinant in \eqref{frs} admits explicit factorized
expressions in the two cases: $r=\infty$, at arbitrary $\alpha$,
and $\alpha=1$, at  arbitrary $r$.

In the first case, $r=\infty$, the entries of the determinant in
\eqref{frs} can be recognized as the moments of the orthogonality
measure of Meixner polynomials $M_j(m;1,\alpha)$; for notation and
summary of properties we refer to \cite{KLS-10}, Section 9.10. Hence,
we can regard the determinant in \eqref{frs} at $r=\infty$ as the Gram
determinant of the Meixner polynomials, equal to $\prod_{j=0}^{s-1}
h_j/\kappa_j^2$, where $h_j$ and $\kappa_j$ are the square norm and
the leading coefficient of the polynomial $M_j(m;1,\alpha)$. Using the
known expressions
\begin{equation}
h_j=\frac{1}{(1-\alpha)\alpha^j},\qquad
\kappa_j=\frac{(\alpha-1)^j}{j!\,\alpha^j},
\end{equation}
we can evaluate the determinant, with the result:
\begin{equation}\label{meixner}
\det_{1\leq j <k\leq s}
\left[\sum_{m=0}^{\infty}m^{j+k-2}\alpha^m\right]
=\prod_{j=0}^{s-1}
\frac{(j!)^2\alpha^{j}}{(1-\alpha)^{2j+1}}.
\end{equation}
This expression allows estimating the large $r$ behavior of the
determinant in \eqref{frs}, at fixed $s$, see \eqref{larger}.

In the second case, $\alpha=1$, the determinant in \eqref{frs} can be
similarly recognized as the Gram determinant of Hahn polynomials
$Q_j(m;0,0,r-1)$, see \cite{KLS-10}, Section 9.5.  The square norm and
the leading coefficient are
\begin{equation}
h_j=\frac{(j+r)!(r-j-1)!}{(2j+1) ((r-1)!)^2},\qquad
\kappa_j=(-1)^j\frac{(2j)!(r-j-1)!}{(j!)^2 (r-1)!}.
\end{equation}
Hence,
\begin{equation}\label{alpha=1b}
\det_{1\leq j <k\leq s} \left[\sum_{m=0}^{r-1}m^{j+k-2}\right]
= \prod_{j=0}^{s-1}
\frac{(j!)^4 (j+r)!}{(2j)!(r-j-1)!(2j+1)!}.
\end{equation}
This leads immediately to expression \eqref{crs} for the quantity
$C_{r,s}$, which, in turn, gives rise to the function $\psi(v)$, see
\eqref{defpsi} and \eqref{psi}.

\section{Resolvents and eigenvalue densities}

Here we explain the origin of \eqref{WI} and
\eqref{WII}, and provide the explicit expression for the eigenvalue
densities associated to the two scenarios appearing in the study of
the random matrix integral \eqref{Irs}.

The explicit solution of the saddle-point equations \eqref{spe1} and
\eqref{spe2} is given by the integral \eqref{Wint} by suitably
specifying the function $U(z)$.  In this way we obtain expressions for
the functions $H_\mathrm{I}(z)$ and $H_\mathrm{II}(z)$, which, when
substituted in \eqref{WHI} and \eqref{WHII}, give rise to \eqref{WI}
and \eqref{WII}, respectively.  The appearing integrals can be
evaluated using
\begin{multline}
\int_a^b\frac{1}{(z-u)\sqrt{(u-a)(b-u)}}\log\frac{u-c}{u-d}\,\rmd u
\\
=
\begin{cases}
\dfrac{2\pi}{\sqrt{(z-a)(z-b)}}
\log\dfrac{\sqrt{a-c}\sqrt{z-b}+\sqrt{b-c}\sqrt{z-a}}
{\sqrt{a-d}\sqrt{z-b}+\sqrt{b-d}\sqrt{z-a}}
&(c,d\leq a)
\\[12pt]
\dfrac{2\pi}{\sqrt{(z-a)(z-b)}}
\log\dfrac{\sqrt{c-a}\sqrt{z-b}+\sqrt{c-b}\sqrt{z-a}}
{\sqrt{d-a}\sqrt{z-b}+\sqrt{d-b}\sqrt{z-a}}
&(c,d\geq b)
\end{cases}
\end{multline}
which holds for $z\in\mathbb{C}\backslash[a,b]$.

Explicit expressions for the eigenvalue densities corresponding to the two
scenarios can be extracted from the resolvents using \eqref{rhodef}.
In the first scenario, the resolvent $W_\mathrm{I}(z)$ has the form
\eqref{WI}, leading to
\begin{equation}
\rho_\mathrm{I}(z)=
\frac{2}{\pi}\arctan\frac{\sqrt{a(b-z)}}{\sqrt{b(z-a)}},\qquad z\in[a,b].
\end{equation}
Recall that $\rho_\mathrm{I}(z)=1$ for $z\in[0,a]$, and  $\rho_\mathrm{I}(z)=0$
for $z\in[b,R]$.

In the second scenario, the resolvent $W_\mathrm{II}(z)$ has the form
\eqref{WII}, and \eqref{rhodef} gives
\begin{equation}
\rho_\mathrm{II}(z)=\frac{2}{\pi}\arctan\frac{\sqrt{a(b-z)}}{\sqrt{b(z-a)}}
-\frac{2}{\pi}\arctan\frac{\sqrt{(R-a)(b-z)}}{\sqrt{(R-b)(z-a)}}+1,
\qquad z\in[a,b]
\end{equation}
with $\rho_\mathrm{II}(z)=1$ for $z\in[0,a] \cup [b,R]$.

As an example, \figurename~\ref{fig-DensityPlots} shows plots of the
eigenvalue densities in the case of $\alpha=0.25$, in which $R_c=3$,
for some $R>3$ and for $R=1.2$, corresponding to the first and the second
scenario, respectively.

\bibliography{thord_bib}
\end{document}